%% file: main.tex
\documentclass[sigconf,nonacm,9pt]{acmart}
\usepackage{array}
\usepackage{booktabs}
\usepackage{graphicx}
\usepackage{xcolor}
\usepackage{pifont}
\usepackage{xurl}
\usepackage{colortbl}
\usepackage{tikz}
\usepackage{flushend}
\definecolor{MarkYes}{HTML}{0B8A6C}
\definecolor{MarkNo}{HTML}{C62828}
\definecolor{MarkPart}{HTML}{C27C0E}
\definecolor{RowOurs}{HTML}{E4F8F2}
\newcommand{\chg}{{\color{MarkYes}\ding{108}}}
\DeclareRobustCommand{\cmark}{{\color{MarkYes}\ding{52}}}
\DeclareRobustCommand{\xmark}{{\color{MarkNo}\ding{56}}}
\DeclareRobustCommand{\pmark}{\tikz[baseline=-0.6ex]{\fill[MarkPart] (0,0) -- (0,0.8ex) arc[start angle=90,end angle=270,radius=0.8ex] -- cycle;\draw[MarkPart,line width=0.5pt] (0,0) circle[radius=0.8ex];}}
\newcommand{\system}{SOMA}
\input{results}

\AtBeginDocument{}
\makeatletter
\@ifundefined{if@ACM@balance}{}{\@ACM@balancefalse}
\@ifclasslater{acmart}{2026/10/31}{%
  \patchcmd{\maketitle}{\setcounter{footnote}{1}}{\setcounter{footnote}{0}}{}{}%
  \patchcmd{\maketitle}{\setcounter{footnote}{1}}{\setcounter{footnote}{0}}{}{}%
}{}
\makeatother
\setcopyright{none}
\begin{document}
\title[\system{}: Transactional Migration of Live Python Sessions]{\system{}: Transactional Migration of Live Python Sessions Across Heterogeneous Runtimes}
\author{Eldar Hasanov}
\authornote{All authors contributed equally to this research.}
\orcid{0009-0008-2157-3867}
\affiliation{\institution{Clusy Inc.}\city{San Francisco}\state{California}\country{USA}}
\email{eldar@clusy.io}
\author{Ju Lin}
\authornotemark[1]
\orcid{0009-0006-7448-9752}
\affiliation{\institution{Clusy Inc.}\city{San Francisco}\state{California}\country{USA}}
\email{ju@clusy.io}
\author{Mohamed Fouzil Ali Syed Ali}
\authornotemark[1]
\orcid{0009-0002-9021-4020}
\affiliation{\institution{Clusy Inc.}\city{San Francisco}\state{California}\country{USA}}
\email{fouzil@clusy.io}
\renewcommand{\shortauthors}{Hasanov et al.}
\begin{abstract}
\input{sections/0-abstract}
\end{abstract}
\keywords{session migration, live state, heterogeneous runtimes, microVMs, containers, GPUs, notebooks, checkpointing}
\maketitle
\hypersetup{pdfcreator={LaTeX},pdfsubject={Python workload migration across heterogeneous runtimes}}
\input{sections/1-introduction}
\input{sections/2-background-and-motivation}
\input{sections/3-design}

\input{sections/5-evaluation}
\input{sections/6-conclusion}
\bibliographystyle{ACM-Reference-Format}
\begingroup
\useOriginalUrlSetting
\newcommand{\showURL}[1]{\unskip\newline\parbox[t]{\linewidth}{\raggedright#1}}
\newcommand{\shownote}[1]{\unskip\newline#1}
\providecommand{\showISBNx}[1]{}
\renewcommand{\showISBNx}[1]{\unskip\newline ISBN~#1.}
\let\SOMAbibHref\href
\renewcommand{\href}[2]{%
  \IfBeginWith{#1}{https://doi.org/}
    {\unskip\newline\SOMAbibHref{#1}{#2}}
    {\SOMAbibHref{#1}{#2}}}
\bibliography{references}
\endgroup
\flushcolsend
\end{document}

%% file: results.tex
\newcommand{\EtwelveText}{We killed the controller at 17 cloud-path crash points: before and after phase-journal writes and between destination creation and journaling.
All 17 restarts recovered one authoritative runtime, preserved acknowledged writes, and matched committed state to the capture cut.
Writes during downtime were refused exactly when admission was closed (8 of 17 cases).
Recovery reclaimed the unjournaled destination and left no runtime leaks.
Median recovery time was 40~s (range: 0--65~s); concurrent requests for the same move created only one destination.}
\newcommand{\EfifteenLiveText}{On the cloud path, an E2B~\cite{e2b2026docs} CPU-to-Modal~\cite{modal2026docs} T4 move passed every check under concurrent writes: all 20 acknowledged writes arrived once, 20 refused writes left no trace, and training continued on the GPU.
The return move failed preflight because the CPU image lacked imported distributions that the prototype could not repair.}
\newcommand{\MoveTimeText}{These measurements use small correctness fixtures.
The CPU-to-T4 move carried a 77.5~KiB compressed capsule: before compression, the serialized namespace was 83.1~KiB and six workspace files totaled 16.1~KiB, plus manifest metadata.
The move took 77~s, with admission closed for 22~s through capture, restoration, validation, and commit.
CPU-to-CPU moves carried 68.1~KiB compressed capsules (62.9~KiB namespaces and four files totaling 16~KiB before compression, plus metadata) and paused execution for 16--26~s.}

%% file: sections/0-abstract.tex
Stateful Python workloads accumulate hours of in-memory state bound to their initial runtime.
When resource needs change or a runtime is reclaimed, continuing elsewhere requires preserving or rebuilding that state.
We present \system{}, a session-oriented migration abstraction for Python workloads that preserves supported application state while transferring execution authority across heterogeneous runtimes, without periodic checkpointing or recomputation.
Its mechanism reconstructs a portable capsule of namespace, files, and environment, preserves object and storage relationships, and commits each move transactionally.
For planned moves, no acknowledged work is lost and an aborted migration leaves the source serving.
The capsule substrate has supported 15K+ Clusy Notebook sessions and 2K+ Python jobs in Portal.
Production spot jobs achieved average compute savings above 41\% relative to equivalent on-demand machines.
Matching hardware to each phase cuts measured job costs by up to 70\%. Sharing a GPU between two sessions raises training utilization from 50\% to 90\%; migration to a larger GPU lets jobs continue after memory exhaustion.

%% file: sections/1-introduction.tex
\section{Introduction}
\label{sec:introduction}

A running Python workload accumulates in-memory objects and files.
That state lives in one process on one machine, yet the machine a workload needs changes as the work proceeds.
An accelerator may be useful for only one phase, or a spot instance may be reclaimed before execution finishes~\cite{aws2026spot}.
For stateful programs, moving live state avoids rebuilding it.
Shared-cluster GPUs are often underutilized~\cite{jeon2019analysis,weng2022mlaas}.

\begin{table*}[t]
\centering
\caption{Comparison of \system{} with prior approaches to moving a live session. \cmark{} yes \quad \pmark{} partial \quad \xmark{} no}
\label{tab:compare}
\small
\renewcommand{\arraystretch}{1.2}
\setlength{\tabcolsep}{4pt}
\newcolumntype{Y}{>{\centering\arraybackslash}m{1.68cm}}
\newcolumntype{Z}{>{\centering\arraybackslash}m{1.72cm}}
\begin{tabular}{>{\raggedright\arraybackslash}m{\dimexpr\textwidth-10.12cm-14\tabcolsep\relax}YYYYZY}
\toprule
\textbf{Approach} & \textbf{Crosses runtime type, OS, ISA} & \textbf{No privileged access} & \textbf{Crosses CPU and GPU models} & \textbf{Live state without app code} & \textbf{Rebuilds \mbox{dependencies}} & \textbf{Atomic handoff} \\
\midrule
VM live migration~\cite{clark2005live} & \xmark & \xmark & \xmark & \cmark & \pmark & \cmark \\
Process checkpointing: CRIU~\cite{criu2026criu}, DMTCP~\cite{ansel2009dmtcp} & \pmark & \pmark & \xmark & \cmark & \pmark & \pmark \\
GPU checkpointing: CRIUgpu~\cite{stoyanov2025criugpu}, PhoenixOS~\cite{wei2025phoenixos} & \xmark & \xmark & \xmark & \cmark & \pmark & \pmark \\
Managed deep-learning job migration~\cite{shukla2022singularity,fluidcr2026} & \xmark & \xmark & \pmark & \cmark & \cmark & \pmark \\
Provider snapshots: Modal~\cite{modal2026snapshots}, E2B~\cite{e2b2026docs}, Firecracker~\cite{agache2020firecracker} & \xmark & \xmark & \xmark & \cmark & \cmark & \pmark \\
Application checkpoints: SkyPilot~\cite{yang2023skypilot} & \cmark & \cmark & \pmark & \xmark & \pmark & \xmark \\
ElasticNotebook~\cite{li2023elasticnotebook} & \cmark & \cmark & \pmark & \cmark & \xmark & \pmark \\
JuPyLive~\cite{attarkhorasani2026jupylive} & \pmark & \cmark & \pmark & \cmark & \cmark & \xmark \\
NotebookOS~\cite{carver2026notebookos} & \xmark & \pmark & \pmark & \cmark & \xmark & \pmark \\
\midrule
\rowcolor{RowOurs} \system{} (this paper) & \cmark & \cmark & \cmark & \cmark & \cmark & \cmark \\
\bottomrule
\end{tabular}
\end{table*}

Python runs in Firecracker~\cite{agache2020firecracker} microVMs, gVisor~\cite{gvisor2026gvisor} sandboxes, Docker~\cite{merkel2014docker} and HPC containers~\cite{kurtzer2017singularity}, scheduled HPC nodes~\cite{yoo2003slurm}, and on laptops.
Existing migration mechanisms differ in the state they preserve and the destinations they support (Table~\ref{tab:compare}).
We therefore make the session durable.

We present \system{}, a session-oriented migration abstraction that preserves supported application state while transferring execution authority across heterogeneous runtimes.
It moves Python workloads between runtimes that can execute Python and satisfy the capsule's portability contract, without a shared execution image or privileged provider interface.
Production deployments include Clusy Notebook~\cite{clusy2026} and Portal~\cite{clusy2026portal}.

Working above the process enables portability but introduces new problems.
Serialization can break shared storage, and replacing a restored parameter can leave its optimizer pointing to the old object.
Concurrent writes and validation code can also alter the state being transferred.
Hardware-induced numerical differences must be distinguished from state loss or corruption.

We contribute \emph{(1)} a portability contract covering stored values, object relationships, and execution authority; \emph{(2)} graph-preserving reconstruction with an atomic, recoverable handoff; and \emph{(3)} an evaluation of controlled migrations and production deployments, including spot workloads.
The experiments exercise cloud microVMs and containers, local processes, two instruction sets, and CPU/GPU transfers.
Five successive training-session moves reproduced all 150 control losses; fault and crash tests checked write preservation and recovery.
Spot case studies examine completion and compute cost under preemption (Section~\ref{sec:portal-spot}).
Code and experiments are at \url{https://github.com/clusy-io/soma}.

%% file: sections/2-background-and-motivation.tex
\section{Motivation and Background}
\label{sec:background}

\paragraph{When a workload must move}
Allocating hardware for a workload's peak demand can leave it idle during other phases: in production notebook traces, reserved GPUs were idle more than 81\% of the time~\cite{carver2026notebookos}.
Conversely, memory exhaustion accounted for 8.8\% of failed deep-learning jobs in one production cluster~\cite{zhang2020empirical,gao2020estimating}, motivating migration to a larger device.
Other triggers include laptop-to-cloud development, maintenance, HPC rescheduling~\cite{schedmd2026preemption}, and hosted-notebook session limits~\cite{google2026colab}.

\paragraph{Spot recovery}
SpotLake~\cite{lee2022spotlake} and \emph{Can't Be Late}~\cite{wu2024cantbelate} show that cost depends jointly on spot price, availability, and interruptions.
Providers reclaim capacity on short notice~\cite{aws2026spot,google2026spot,azure2026spot}, with notice periods varying by cloud.
SkyPilot~\cite{yang2023skypilot} places jobs across clouds and restores application checkpoints on compatible machines.

\paragraph{What existing mechanisms preserve}
Table~\ref{tab:compare} compares six migration properties.
VM live migration~\cite{clark2005live} transfers an execution image and introduced the view of migration as a transaction between two hosts, which we adopt for sessions.
Process checkpointing preserves native state, including threads and open handles, but constrains the destination: CRIU~\cite{criu2026criu} and DMTCP~\cite{ansel2009dmtcp} restore Linux processes on the same instruction set.
GPU checkpointing~\cite{stoyanov2025criugpu} adds device constraints; NVIDIA's native restore~\cite{nvidia2026cudacheckpoint} requires the same chip type and sufficient memory.
Managed training systems interpose on device operations within a fleet~\cite{xiao2018gandiva,fluidcr2026}.
Singularity~\cite{shukla2022singularity} requires the GPU type to remain fixed.
Provider services impose their own restore contracts: Modal~\cite{modal2026snapshots} offers experimental GPU memory snapshots tied to the original worker type.

\paragraph{Application-level recovery}
Application checkpoints~\cite{mohan2021checkfreq,eisenman2022checknrun,wang2023gemini} can cross provider and hardware boundaries when the application saves all required state and destination software is compatible.
Periodic checkpointing can add overhead and interfere with the main training workload~\cite{hasanov2026checkwork}.
Multi-cloud managers use checkpoints to support placement decisions~\cite{yang2023skypilot,stoica2021sky}.
ElasticNotebook~\cite{li2023elasticnotebook} chooses between copying and recomputing each variable and reconstructs objects that cannot be serialized.
JuPyLive~\cite{attarkhorasani2026jupylive} extends it to laptop-to-HPC migration, but ignores notebook edits made while migration waits.
NotebookOS~\cite{carver2026notebookos} allocates GPUs on demand by replicating kernels within its platform; Kishu~\cite{li2025kishu} versions session state.

%% file: sections/3-design.tex
\section{Design}
\label{sec:design}

\begin{figure}[t]
\centering
\includegraphics[width=0.95\columnwidth]{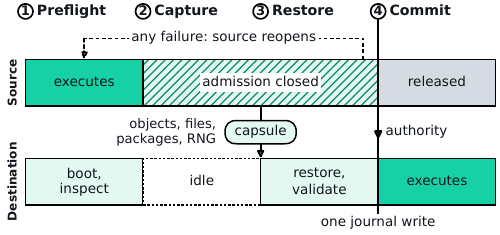}
\caption{Validate, commit, then release the source.}
\Description{Two horizontal lanes, source and destination runtime, across four numbered phases: preflight, capture, restore, commit.
The source executes admitted work, then closes admission from capture until commit; the capsule of objects, files, packages and RNG state drops from source to destination at the capture/restore boundary.
The destination boots and inspects, idles, then restores and validates. At commit, one journal write moves authority down to the destination, which executes while the source is released.
A dashed path marks that any failure before commit reopens the source.}
\label{fig:protocol}
\end{figure}

\system{} separates a persistent logical session from the runtime that executes it.
A \emph{session} is a Python execution together with its supported state; a \emph{runtime} can start Python and execute code on request.
The controller needs three operations, create, execute, and delete, and moves state via the execute channel and an object store.
Figure~\ref{fig:protocol} shows a planned move.

\subsection{A Capsule with a Portability Contract}
\label{sec:contract}

A move is correct when, at takeover, the destination holds the source state at capture, except for declared adaptations such as moving CUDA tensors to a CPU.
The contract covers stored values, object relationships, and continuation state, including random generators.
Stored bytes must match exactly; recomputed floating-point values are checked separately under a stated tolerance~\cite{pytorch2026randomness}.
Accepted execution completes before capture or at the committed destination; rejected execution does no work.
This guarantee covers execution routed through the controller.

The capsule contains the supported namespace serialized with \texttt{dill}~\cite{mckerns2011building}, workspace files with modes and hashes, environment requirements, and reconstruction metadata.
It also records the storage relationships that must survive, supported gradients, random-generator states, and fingerprints of the state at the capture cut.
Files and packages are restored before the namespace because an object may refer to a path or an installed class.
The capsule records source package versions; restoration reinstalls these dependencies while retaining the destination's accelerator stack, including PyTorch~\cite{paszke2019pytorch}, Triton~\cite{tillet2019triton}, and NVIDIA libraries.
This avoids inheriting CUDA packages built for the source device.
If loading reveals another missing module, \system{} installs its recorded version and retries.
An unavailable distribution can still prevent the move, as the cloud return-path test in Section~\ref{sec:evaluation} demonstrates.

\paragraph{Preflight and state coverage}
For a planned move, preflight inspects the provisioned destination's Python and serializer versions, devices, and installed packages.
The manifest classifies objects as supported, transformed, rejected, or unknown.
Unsafe objects, irreparable relationships, and unavailable dependencies block capture; strict policy also rejects unknown state and process effects.
The contract excludes instruction-level continuity and live threads or subprocesses.
It also excludes sockets, OS handles, generic autograd graphs, distributed communicators, and external-service rollback.
Preflight predicts whether reconstruction is feasible; checks after restoration test the resulting state.
A successful environment check therefore does not replace validation of the restored objects and their relationships.
The manifest reports exclusions and adaptations; unknown state is never reported as verified.
Restoration checks the trusted capsule's size and SHA-256 before loading.

\subsection{Preserving the Object Graph}

\paragraph{Class identity}
PyTorch~\cite{paszke2019pytorch} can detach optimizer submodules from their packages.
Serializing their classes by value can break type checks after restoration.
Capture briefly reattaches the submodules so installed classes serialize by reference.

\paragraph{Storage repair must preserve object identity}
A tensor view is a distinct object sharing its base's storage.
PyTorch~\cite{pytorch2026serialization} preserves views in its own serialization, but namespace reconstruction must also preserve references from other objects.
The capsule records each supported view's anchor, shape, stride, offset, type, and reference paths, including anchors without namespace bindings.
For a leaf tensor, the adapter repoints storage without replacing the object.
Module registration, optimizer parameter groups and state keys, repeated references, and \texttt{requires\_grad} therefore retain the same object identity.
For example, replacing a model parameter with an equal-valued tensor would leave an optimizer holding the old parameter; a value-only check could miss that error.
A view is rebuilt from its anchor so its storage relationship survives.
This also applies to supported gradient-bearing views, while generic autograd graphs remain outside the contract.

NumPy~\cite{harris2020array} uses buffers that cannot be repointed.
The adapter rebinds all covered references and refuses restoration if an uncovered holder retains the stale array.
Failed repairs block commit.

\paragraph{Device and arithmetic differences}
CUDA storage maps to CPU when the destination lacks a compatible device; CPU objects remain there until the application moves them.
Restoration reinstates Python, NumPy~\cite{harris2020array}, and PyTorch~\cite{paszke2019pytorch} CPU generator states, plus CUDA generator states when both endpoints support them.
CPU destinations declare CUDA streams dropped.
TF32, math libraries, and SIMD implementations can change recomputed values~\cite{nvidia2020a100,pytorch2026randomness}.
Generator states and integer-derived draws must match exactly; library-dependent draws are reported separately.

\subsection{Consistent Cut and Atomic Handoff}

\paragraph{Admission closes before capture}
One journal transaction either resolves the authoritative runtime and registers an execution as in flight, or refuses it because admission is closed.
Closing admission uses the same transaction.
The notebook path drains admitted requests and establishes a kernel barrier; the job path pauses at a safepoint (Section~\ref{sec:implementation}).
The gate belongs to the runtime: requests through earlier move identifiers follow the current authority, and concurrent outgoing moves are refused.
The capture boundary separates completed requests from work that may execute after the handoff.
The destination remains unavailable to callers while its state is reconstructed and checked; merely creating it does not transfer authority.
This gives one execution authority a consistent capture boundary~\cite{chandy1985distributed}.

\paragraph{Validation is not user work}
Capture computes reference digests without running the user's model.
The destination checks parameters, optimizers, generators, required views, and workspace hashes against them.
Checks that execute the model use an isolated child process, or a disposable copy where CUDA initialization prevents isolation.
Neither technique alone contains every effect: a child process shares the file system, and a copied model can keep a hook that references the original namespace.
The controller therefore fingerprints the supported state and workspace before and after validation and aborts on any change or on a comparison it could not make.
Before commit it rechecks the destination against the capture cut.
Checks that cannot run safely remain unverified.

\paragraph{Commit moves authority before the source is released}
A single journal update makes the destination authoritative and reopens admission, following a two-phase commit structure~\cite{gray1978notes}.
Before commit, a failed planned move discards the destination and reopens the source; afterward, recovery completes the handoff and releases the source.
If preemption has already reclaimed the source, recovery retains its durable capsule until a destination is ready.
Persisted phases let a restarted controller finish restoration and validation before exposing the destination.
Move names identify orphan destinations created before their identifiers were journaled.
Local SQLite~\cite{sqlite2026about} transactions support controller recovery.

\input{sections/4-implementation}

%% file: sections/4-implementation.tex
\subsection{Implementation and Deployment}
\label{sec:implementation}

Clusy Notebook~\cite{clusy2026} lets users and agents move interactive sessions among cloud and local runtimes.
CPU sessions default to E2B~\cite{e2b2026docs} sandboxes backed by Firecracker~\cite{agache2020firecracker}; GPU sessions use Modal~\cite{modal2026docs,modal2026security} containers isolated by gVisor~\cite{gvisor2026gvisor}.
A CPU-to-GPU move changes provider, isolation technology, device, and base image.
Workspace files transfer through presigned object-store URLs with their metadata.
This deployment captures before source destruction and restores lazily on the next execution.
Adapters for local processes and Docker~\cite{merkel2014docker} support cross-OS/ISA tests.

\paragraph{Portal job execution}
Portal~\cite{clusy2026portal} instruments supported Python loops with safepoints and retains their positions and paused locals in the capsule.
This continuation record describes Python application state; it does not contain native stacks or child processes.
Applications that launch external workers must arrange for those workers to restart, as in the forecasting case study.
Users or agents submit scripts through its CLI or API and select a compatible destination for resumption.
On a spot notice, the node pauses execution and uploads the capsule to durable intermediate storage in the source provider and region.
After persistence, it publishes the manifest and provisions a destination to restore the capsule.
The pause and durable save must finish within the notice period; provisioning and restoration can follow after the source is reclaimed.
This avoids periodic saves for supported jobs.

%% file: sections/5-evaluation.tex
\section{Evaluation}
\label{sec:evaluation}

We evaluate fidelity, handoff recovery, latency, and cost.

\subsection{Setup}

Table~\ref{tab:crossings} lists migration routes: the first two test capsule reconstruction; the others include transactional handoff.
Cloud tests use the notebook deployment (Section~\ref{sec:implementation}) with 8-vCPU x86-64 CPU microVMs and T4 or A100-40GB GPU containers; timings include wide-area round trips.
Offline tests use a macOS arm64 host, a Linux arm64 container with the production sandbox image, and an emulated x86-64 container.
Controlled tests use Python 3.11, \texttt{dill}~\cite{mckerns2011building} 0.3.9, and PyTorch~\cite{paszke2019pytorch} 2.7.1.

We test a ResNet-18~\cite{he2016resnet} training session with synthetic $3\times32\times32$ inputs, optimizer, scheduler, loader, generators, and preprocessing artifacts; an extended fixture with shared parameters, tensor and NumPy~\cite{harris2020array} views, gradients, and files; and an array-based training session with views, aliases, a cycle, and a log file.

\begin{table}[t]
\centering
\caption{Every route kept the session exactly or failed safely, over 50 trials each. \chg{} marks what the route changes.}
\label{tab:crossings}
\footnotesize
\renewcommand{\arraystretch}{1.15}
\setlength{\tabcolsep}{2.6pt}
\newcommand{\rh}[1]{\rotatebox{90}{\textbf{#1}}}
\begin{tabular}{@{}lcccccl@{}}
\toprule
\textbf{Route} & \rh{Provider} & \rh{Isolation} & \rh{OS} & \rh{ISA} & \rh{Device} & \textbf{Result} \\
\midrule
T4 $\rightarrow$ T4 & & & & & & training bitwise identical \\
T4 $\leftrightarrow$ CPU $\leftrightarrow$ A100 & \chg & \chg & & & \chg & stored state exact \\
CPU $\rightarrow$ T4, concurrent writes & \chg & \chg & & & \chg & exact, no write lost \\
CPU $\rightarrow$ CPU, injected faults & & & & & & aborted, source kept \\
CPU $\rightarrow$ CPU, controller crashes & & & & & & recovered, no write lost \\
macOS process $\leftrightarrow$ Linux container & & \chg & \chg & & & exact, no write lost \\
arm64 $\leftrightarrow$ x86-64 container & & & & \chg & & stored state exact \\
\bottomrule
\end{tabular}
\end{table}

\subsection{State Fidelity}

\paragraph{Exact continuation}
We destroyed and reprovisioned a T4 container across five training-session moves, running 30 steps between moves.
All 150 losses matched two independent uninterrupted controls bitwise.
Generator states (including CUDA) and parameter, optimizer, scheduler, and loader-generator digests also matched.

\paragraph{Across devices and GPU models}
A nine-move chain crossed T4, CPU, and A100 runtimes.
Every move preserved parameter bytes, tied-weight identity, and optimizer step counts.
Generator states were preserved for Python, NumPy~\cite{harris2020array}, PyTorch~\cite{paszke2019pytorch} on CPU, and named generators.
CUDA generator states matched on the three GPU-to-GPU moves, including T4--A100 transfers, and were declared dropped on the three moves into CPU runtimes.

\paragraph{Outgrowing a GPU}
We increased two training jobs' memory demand until a CUDA out-of-memory error triggered migration to a larger GPU.
Fine-tuning the small GPT-2~\cite{radford2019gpt2} model on WikiText-2~\cite{merity2017pointer} at sequence length 1024 required 16.2~GiB, exceeding T4 memory.
Runs moved to an A100 after 20, 120, and 1080 completed steps and finished there.
Across 10 tests, ViT-B/16~\cite{dosovitskiy2021vit} training on CIFAR-10~\cite{krizhevsky2009cifar} with larger images exhausted T4 and then L4 memory, requiring T4--L4--A100 migration within a session.

\paragraph{Across operating systems and instruction sets}
The combined fixture passed all 101 checks on a macOS-host-to-Linux-container round trip, covering views, optimizer state, gradients, generators, files, and routed writes.
A NumPy~\cite{harris2020array} session also survived an arm64--x86-64 round trip with exact stored state, shared views, optimizer references, aliases, and a cycle intact.
The first macOS-to-Linux attempt aborted because Gaussian draws differed in their last bit between Apple's math library and glibc despite identical generator state.
The check compares generator state exactly and reports library-dependent draws separately.

\begin{figure}[t]
\centering
\includegraphics[width=0.95\columnwidth]{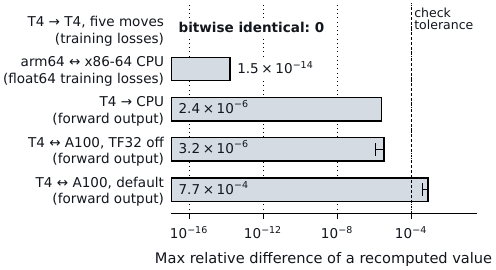}
\caption{Stored state is exact; recomputation can vary.}
\Description{Horizontal bars on a logarithmic axis: T4 to T4 bitwise identical; arm64 to x86-64 1.5e-14; T4 to CPU 2.4e-6; T4 and A100 with TF32 off 3.2e-6; T4 and A100 default 7.7e-4, beyond a 1e-4 tolerance line.}
\label{fig:fidelity}
\end{figure}

\paragraph{Recomputed arithmetic}
A forward pass after T4--A100 migration deviated by up to $7.7\times10^{-4}$, exceeding the $10^{-4}$ tolerance, but by at most $3.2\times10^{-6}$ with TF32 disabled on the same restored bytes (Figure~\ref{fig:fidelity}).
Float64 losses matched the arm64 control bitwise until the first x86-64 step, then agreed within $1.5\times10^{-14}$.

\subsection{Handoff and Recovery}

\paragraph{Concurrent writes}
We tested three moves between a host process and a container while a writer issued numbered updates.
It sent counter increments and file appends through the controller, including requests addressed to an earlier move identifier.
The first move committed, the second aborted on a corrupted capsule, and the third recovered after a controller crash following capture.
Exactly one runtime remained authoritative.
All 63 acknowledged writes appeared exactly once at the final runtime; all 73 refused writes, including three during controller downtime, left no trace.
The aborted move left the source unchanged.
\EfifteenLiveText{}

\paragraph{Injected faults}
The cloud path completed a clean move and handled six injected faults: a missing module, allocation failure, and corrupted capsule aborted after restoration; an unavailable distribution blocked preflight; a nested open log aborted transfer; and an excluded open file handle remained at the source while the move committed.
Aborts left the source authoritative and serving.
\paragraph{Crash recovery}
\EtwelveText{}

\paragraph{Detecting regressions}
Reintroducing three past defects triggered the expected failed checks: validation stepping the real optimizer, an execution route bypassing admission, and storage repair replacing a parameter.
These cases test whether the checks detect a violated contract, beyond accepting successful moves.
Regression coverage includes shared parameters, views, state-mutating hooks, and admission races.

\subsection{Migration Latency}

\MoveTimeText{}

The pause includes draining admitted work, capture, restore, validation, and commit.
Total move time also includes destination setup and source release.
In the CPU-to-T4 run, setup and preflight took 25.1~s, draining 3.4~s, capture 2.7~s, and restoration 6.2~s.
Validation and the boundary recheck took 9.9~s; the journal commit took less than a millisecond.
Source release took a further 27.6~s after authority had transferred, so it extended total move time without extending the execution pause.
Controller overhead accounted for the remaining 2.1~s.

Reported sizes measure compressed archives, rather than memory footprints or wire traffic.
These small correctness fixtures expose the cost of the handoff protocol, including remote round trips and checks.
They do not establish transfer throughput for large capsules; the CPU-to-T4 timing should not be extrapolated as a bytes-per-second rate.

In a separate controlled benchmark of the deployed switch path, batching file persistence cut median switch-request time from 302 to 96~s for 64 files totaling 8~MiB, with two trials per build.
That timing includes both workspace persistence passes.

\subsection{Compute Cost and Utilization}

\begin{figure}[t]
\centering
\includegraphics[width=0.95\columnwidth]{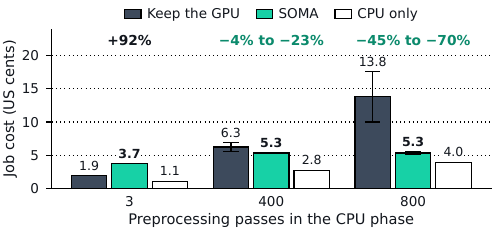}
\caption{Migration saves 45--70\% versus GPU-only use.}
\Description{Grouped vertical bars for three CPU-phase lengths (3, 400, and 800 preprocessing passes) comparing keeping the GPU, SOMA, and CPU only, with SOMA's change against keeping the GPU above each group: +92\%, $-$4\% to $-$23\%, and $-$45\% to $-$70\%.
Short phase: 1.9, 3.7, and 1.1 cents; 400 passes: 6.3, 5.3, and 2.8 cents; 800 passes: 13.8, 5.3, and 4.0 cents.}
\label{fig:cost}
\end{figure}

\paragraph{A measured job}
A job prepares data on a CPU, trains on a T4, and analyzes on a CPU, carrying preprocessing artifacts that the analysis consumes.
Figure~\ref{fig:cost} compares costs at list prices across repeated runs with a retained GPU, migration, or CPU-only execution.
The migration comparison charges both runtimes while their allocations overlap.
Keeping an idle source available during capture therefore has a cost, even though it is no longer performing useful computation.
With a long CPU phase, \system{} costs 45\% to 70\% less than keeping the GPU.
Part of the gain came from preparation using two PyTorch~\cite{paszke2019pytorch} threads in the GPU container versus four in the CPU microVM: 384--613~s versus 131--134~s.
With a short CPU phase, the move costs more than it saves, and CPU-only execution is cheapest because the benchmark's training is small.
Restarting on the GPU repeated 57~s of preparation and matched migration cost even for the short job.
A complete application checkpoint reproduced the final parameters and analysis but required 30 lines of application code and cost 2.7$\times$ as much as migration, with both runtimes active during save, transfer, and load.
Without loader and preparation state, training inputs changed and analysis failed.

\begin{figure}[ht]
\centering
\includegraphics[width=0.95\columnwidth]{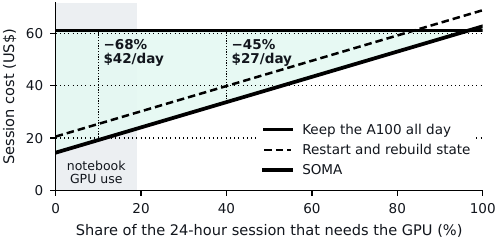}
\caption{Projected savings grow as GPU use falls.}
\Description{Session cost over 24 hours versus the share of time that needs the GPU. Keeping the A100 all day is flat at about 61 dollars; SOMA rises from about 15 dollars and meets it near 95 percent; restart-and-rebuild lies above SOMA. Annotations mark savings of 68 percent or 42 dollars a day at 10 percent GPU time and 45 percent or 27 dollars a day at 40 percent. A shaded band covers 0 to 19 percent GPU time, representing observed notebook GPU use.}
\label{fig:modelday}
\end{figure}

\paragraph{24-hour projection from measured timings}
In an Adobe notebook trace, about three quarters of sessions used their GPU for at most 5\% of their lifetime~\cite{carver2026notebookos}.
Figure~\ref{fig:modelday} projects costs for a 24-hour session with eight GPU episodes, charging each move 120~s at both runtimes' rates (measured median: 66~s).
The restart baseline rebuilds state for 15 minutes per move, adding four hours of work.
Against \$61 per day for a retained A100, migration saves 57\% (\$35) at 25\% GPU use and 68\% (\$42) at 10\% use.
It becomes more expensive above roughly 97\% GPU use.
A smaller GPU premium limits projected T4 savings to 31\%.

\begin{figure}[t]
\centering
\includegraphics[width=0.95\columnwidth]{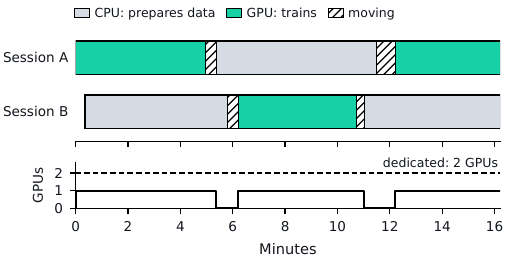}
\caption{GPU sharing raises training utilization: 50\% to 90\%.}
\Description{Two lanes, session A and session B, over 16 minutes. A trains on the GPU while B prepares data on a CPU; they trade runtimes twice, each move shown as a short hatched block. A lower strip counts the GPU runtimes alive: never more than one, against a dashed line at two for dedicated GPUs.}
\label{fig:share}
\end{figure}

\paragraph{Two sessions, one GPU}
Active learning alternates pool scoring and retraining~\cite{settles2009active,sener2018coreset}; CPU data preparation likewise leaves accelerators idle~\cite{mohan2021datastalls}.
Two such sessions, offset by half a cycle, can exchange one CPU and one GPU runtime at phase boundaries, extending cluster-level GPU time-sharing~\cite{xiao2018gandiva} to live sessions.
We tested four-minute phases of NumPy~\cite{harris2020array} scoring and CNN training on an A100 (Figure~\ref{fig:share}).
All four moves preserved parameters and step counters; a swap, including both moves and state checks, took 104~s.
Only one GPU runtime existed at a time.
Over the completed cycle the shared A100 trained 90\% of the time it was held, where dedicated GPUs would train half the time, and the pair cost 31\% less than the same phases on two dedicated A100s at list prices.
With 30-minute phases over eight hours, the model projects 36\% savings (about \$15 per pair), 8.4 rather than 16 GPU-hours, and 5\% longer completion time.
Projected savings on T4s are 13\%.

\input{sections/5a-portal-spot}

\paragraph{Operational limits}
Users, agents, or rules request a move.
The migration mechanism preserves state and transfers authority; a separate policy selects a destination and decides when to resume.
The destination needs sufficient memory and must recreate CUDA streams and compiled kernels, subject to the exclusions in Section~\ref{sec:contract}.
Long native calls or short preemption notices can prevent capture; capacity shortages can delay provisioning and resumption.
Recovery from preemption requires a durable capsule before the source disappears.
Abrupt loss before that save is outside this guarantee.
The journal must remain available to the recovering controller.
Our crash experiments restart the controller process with its journal intact; they do not cover loss of coordinator storage or coordination across network partitions.

\paragraph{Evaluation scope}
Deployment counts and controlled trials describe different populations.
The notebook and Portal totals measure usage of the capsule substrate; they are not counts of successful transactional handoffs.
The fault tests exercise particular protocol boundaries, while the spot study reports recovery and cost for retained jobs.
Quantifying production failure rates would require per-attempt outcomes, notice durations, capsule sizes, and resumption delays, including failed or superseded attempts.

%% file: sections/5a-portal-spot.tex
\subsection{Python Jobs on Spot Compute}
\label{sec:portal-spot}

Portal has run 2K+ Python jobs alongside 15K+ Clusy Notebook sessions.
Reported spot jobs saved over 41\% on average versus equivalent on-demand machines, with no jobs lost to preemption in the reported cohort.

\begin{figure}[t]
\centering
\includegraphics[width=\columnwidth]{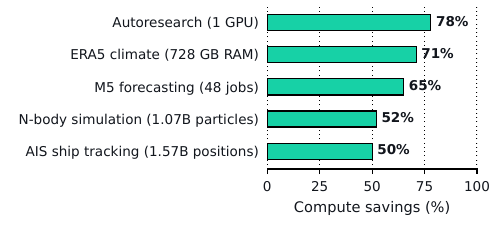}
\caption{Spot costs 50--78\% less than AWS on demand.}
\Description{Five green horizontal bars show compute savings against the AWS on-demand baseline identified in the caption. Autoresearch, one GPU, saves 78 percent; ERA5 climate analysis, 728 GB in memory, 71 percent; M5 forecasting, 48 jobs, 65 percent; N-body simulation, 1.07 billion particles, 52 percent; and AIS ship tracking, 1.57 billion positions, 50 percent. Black labels show workload names, scales, and savings.}
\label{fig:portal-cases}
\end{figure}

Figure~\ref{fig:portal-cases} compares five spot workloads with AWS on-demand pricing~\cite{clusy2026portalcases}, using records checked November 4--5, 2026.
Autoresearch~\cite{karpathy2026autoresearch} ran on Azure; ERA5~\cite{hersbach2020era5} analysis, forecasting, $N$-body simulation, and AIS analysis ran on AWS.
AWS costs use observed machine time; the reported Azure comparison also includes cross-provider price differences.
The ERA5 case analyzes 20 years of hourly climate data with 728~GB in memory.
The simulation uses 1.07 billion particles on a $2048^3$ mesh; ship tracking processes 1.57 billion positions from about 80,000 vessels.
The figure reports selected completed workloads.

\paragraph{M5~\cite{makridakis2022m5} forecasting}
We forecast 30,490 product--store series at 53 origins.
The models include statistical methods, LightGBM~\cite{ke2017lightgbm}, and Moirai 2.0~\cite{liu2025moirai}.
The 54 retained jobs completed in 3.28 hours and used 75.0 machine-hours across nine AWS, Google Cloud, and Azure regions.
Twenty-five jobs underwent 69 notice-triggered resumptions (68 AWS, one Google Cloud), with up to eight resumptions per job.
Replacement machines came from other instance pools within the original cloud.
Recovery required application support.
Workers saved completed units, indexed by store and forecast origin, before stopping on pause.
They restarted after restoration.
Completed units were retained, while unfinished units could repeat.
Pricing measured machine time gives \$43.09 versus \$124.25 at the providers' on-demand rates, a 65.3\% reduction.
Figure~\ref{fig:portal-cases} uses the 48 AWS jobs from this cohort so its comparator remains AWS on demand.
Costs cover retained compute allocations and exclude storage, transfers, and superseded launches.
Savings reflect spot pricing without credit for avoided recomputation.

%% file: sections/6-conclusion.tex
\section{Conclusion}
\label{sec:conclusion}

\system{} turns fixed hardware allocation into a per-phase decision, reducing cost when a GPU is needed only part of the time.
Portal uses the same mechanism to resume jobs on replacement spot capacity.
Future work will broaden state coverage, evaluate placement policies, and explore replicated coordination and incremental capsules.
The results show how a persistent session can outlive one runtime while retaining an explicit boundary for unsupported state.
Our goal is to make Python runtimes as replaceable as the infrastructure beneath stateless services.